\documentstyle[prl,aps,twocolumn]{revtex}
\begin{document}

\title{  Muon Spin Relaxation Investigation of the 
Spin Dynamics of Geometrically Frustrated
Antiferromagnets  Y$_2$Mo$_2$O$_7$ and Tb$_2$Mo$_2$O$_7$  \\}

\author{S. R. Dunsiger${^1}$,
 R. F. Kiefl$^{1,2}$, K. H. Chow${^3}$, B. D. 
Gaulin${^4}$, M. J. P. Gingras$^{5,*}$,
 J. E. Greedan${^6}$, A. Keren${^7}$,
K. Kojima${^7}$, G. M. Luke${^7}$,
W. A. MacFarlane${^1}$, N. P. Raju${^6}$, J. E. Sonier${^1}$,
 Y. J. Uemura${^7}$, W. D. Wu${^7}$ }

\address{$^1$ Department of Physics and Astronomy,
 University of British Columbia,
 V6T-1Z1, Canada\\
$^2$ Canadian Institute for Advanced Research, 
University of British Columbia, V6T-1Z1, Canada\\
$^3$ Clarendon Laboratory, Oxford University, Parks Road, Oxford, OX1 3PU,
 U.K.\\
$^4$ Department of Physics and Astronomy,  McMaster University,
 Hamilton, Ontario,
L8S 4M1, Canada\\
$^{5,*}$ TRIUMF, 4004 Wesbrook Mall, Vancouver, V6T 2A3, Canada\\
$^6$ Department of Chemistry, McMaster University, Hamilton, Ontario,
L8S 4M1, Canada\\
 $^7$ Department of Physics, Columbia University, New York, NY 10027, USA}

\date{\today}
\maketitle

\begin{abstract}

\noindent The spin dynamics of  geometrically frustrated pyrochlore
antiferromagnets Y$_2$Mo$_2$O$_7$ and Tb$_2$Mo$_2$O$_7$
have been investigated
using muon spin relaxation. A dramatic slowing
down of the moment  fluctuations occurs as one approaches 
the spin freezing temperatures (T$_F$=22 K and 25 K respectively) 
from above.
Below T$_F$ there is a  disordered magnetic state similar to that 
found in a  spin glass but with  a  
residual muon spin relaxation rate at low temperatures.
These results show that there is a large density of states for 
magnetic excitations in these systems near zero energy. 
 
\vspace{1cm}
\noindent $^*$Address after September 01, 1996: Dept. of Physics, University of
Waterloo, Waterloo, Ontario, N2L-3G1, CANADA

\end{abstract}

%\newpage

\noindent Antiferromagnets which are frustrated or diluted can  
exhibit novel   electronic and magnetic behaviour. 
Recently, there has been considerable interest in the behaviour of
systems where the
natural antiferromagnetic coupling between ions is frustrated
by the geometry of the lattice. In two dimensions,
Heisenberg spins  on  triangular and   corner
sharing triangular (kagom\'{e}) lattices are simple examples of
geometric frustration, while in three dimensions, the most well
studied systems have a  pyrochlore structure, in which the
magnetic ions occupy a lattice of corner sharing tetrahedra.
A system of Heisenberg spins interacting via nearest-neighbor
antiferromagnetic couplings on the pyrochlore lattice displays a
classical ground state with macroscopic degeneracy, since the lowest
energy spin configuration requires only that $\Sigma_{i=1}^{4} {\bf
S}_i=0$ for each tetrahedron.  This feature led Villain to argue that
these systems remain in a {\it cooperative paramagnetic} state with
only short range spin-spin correlations for all T$>$0~\cite{villain}
and this has been confirmed by Monte Carlo
simulations~\cite{reimers92}.  Possibly, the most interesting feature
of the ground state of pyrochlore~\cite{reimers91} and 
kagom\'e~\cite{harris,chalker} lattice antiferromagnets is the prediction
 of a dispersionless spin-wave branch (``zero modes'').  
These zero modes manifestly affect the thermodynamics
 of these classical systems as
demonstrated by Monte Carlo simulations, where the low temperature 
specific heat, $C_v$,
falls below the classical value $k_B$ expected from equipartition 
of energy~\cite{reimers92,chalker,reimers_berlinsky}.  Also, again
because of these zero modes, the kagom\'e and pyrochlore antiferromagnets
display large spin fluctuations down to T$=0^+$~\cite{keren}.  
However, further nearest neighbour exchange~\cite{reimers91},
 magnetic anisotropy
and fluctuations~\cite{obd} may lift this  classical ground state
degeneracy.

A wide variety of interesting magnetic behaviour has been observed
in real systems.
Neutron scattering results on 
FeF$_3$~\cite{fef3} show a
transition to a non-collinear long range ordered (LRO) state, 
in which the spins on a tetrahedron point away from 
the center.  
However, a large number of oxide pyrochlores do not show 
N\'eel LRO.
Bulk magnetic susceptibility measurements\cite{raju} on the
pyrochlore  Y$_2$Mo$_2$O$_7$  show strong irreversible behaviour 
below T$_F$=22 K, characteristic of
 spin glass ordering, even though the level of disorder is immeasurably 
small. 
Recent measurements of the dc magnetization of Y$_2$Mo$_2$O$_7$
 show a divergent 
non-linear susceptibility
 at T$\approx 22$ K~\cite{nonlinsus}, which is a signature of 
a true thermodynamic spin glass phase transition.
Inelastic
neutron scattering data~\cite{neufluct} on Tb$_2$Mo$_2$O$_7$
 confirm there is rapid
slowing down of the Tb spins 
  as one approaches
T$_F$=25 K from above and the absence of LRO  below T$_F$.
The observed  strong
diffuse scattering in Tb$_2$Mo$_2$O$_7$
 indicates the presence of  short 
range correlations  between the moments, which are frozen on a 
time scale of about $10^{-11}$ s. 

In this letter we report an  investigation of the low
temperature magnetic properties of pyrochlores Y$_2$Mo$_2$O$_7$ and
Tb$_2$Mo$_2$O$_7$ using the technique of muon spin rotation/relaxation 
($\mu$SR ),
which is sensitive to spin fluctuation rates 
in the range
$10^{4}-10^{11} $ s$^{-1}$~\cite{musr}, below that detectable with 
neutron scattering.
We find that, despite its  nominally disorder free structure,
the magnetic behaviour in Y$_2$Mo$_2$O$_7$ is close to that
observed in conventional random spin glasses.   Specifically, 
 a large  static
internal magnetic field with a very broad
distribution develops below T$_F$,
 such that no coherent muon spin precession is
observed. At the same time, the muon spin
relaxation rate  $1/T_1$  decreases according to a power law 
 with decreasing temperature.
A similar magnetic transition occurs in Tb$_2$Mo$_2$O$_7$.
The most remarkable  feature in the data is 
the presence of  a sizeable residual spin relaxation rate 
at low temperatures, which is not  evident from previous data on
conventional metallic spin
glasses like CuMn~\cite{tomo}, AuMn~\cite{pinkvos} and 
amorphous-FeMn~\cite{femn}.
This is direct evidence  for
a larger  density of magnetic excitations near zero energy than in conventional
random spin glasses.

Details on the  preparation of the Y$_2$Mo$_2$O$_7$ and 
Tb$_2$Mo$_2$O$_7$  samples
are given elsewhere~\cite{raju}. 
Pyrochlores crystallize with an fcc structure containing eight formula units
per conventional unit cell and space group Fd$\bar{3}$m. The ions on the 16d 
site form a network of corner sharing tetrahedra;
the 16c sites constitute an identical sublattice,
 displaced by ($\frac{1}{2},\frac{1}{2},\frac{1}{2}$). 
 Mo$^{4+}$ ions occupy the
16c site,  Y$^{3+}$ or Tb$^{3+}$ ions the 16d site. The Tb$^{3+}$ ion 
 has a large
 magnetic moment of $\sim 9\mu_{B}$ ,  roughly nine
times larger than  that of Mo$^{4+}$, whereas  Y$^{3+}$ is diamagnetic.
Y$_2$Mo$_2$O$_7$ and Tb$_2$Mo$_2$O$_7$ are  semiconductors with small 
band gaps of 0.013 and
0.007 eV respectively~\cite{bandgap}.
The samples  in this study were characterized by
magnetic susceptibility and x-ray diffraction.
Sharp irreversibilities in the magnetization were
observed at  spin freezing temperatures of 22 K and 25 K respectively, 
consistent with that seen in other highly
stoichiometric samples of Y$_2$Mo$_2$O$_7$ and Tb$_2$Mo$_2$O$_7$. 
>From  Rietveld profile refinements
 of neutron diffraction 
measurements~\cite{neufluct}, one can say the concentration of 
oxygen vacancies,  likely the main source of 
crystalline disorder in these materials,  is below the detectable limit of 
$1\%$. 
Since the ionic radii of Y$^{3+}$ and Mo$^{4+}$, as well as that of 
Tb$^{3+}$ and Mo$^{4+}$ are very different,
there should be no admixing between the 16c and 16d cations.
  This is confirmed by analysis of X-ray data. 

$\mu$SR measurements  were made at TRIUMF in a $^4$He gas flow cryostat 
for temperatures 
above 2 K and in an 
 Oxford Intruments Model 400 top loading 
dilution refrigerator (DR) for lower temperatures. 
For the DR measurements the pressed polycrystalline 
pellets were  varnished onto
an Ag plate and covered in thin Ag foil, which was bolted
to the cold finger.
In a $\mu$SR experiment the observed quantity  is the time
evolution of the muon spin polarization, which depends on the
distribution of internal magnetic fields  and their temporal  fluctuations.
 In a longitudinal field
(LF) geometry an external magnetic field is directed along
the initial polarization direction. 
The present  measurements were made in  a small longitudinal
field   to quench any spin relaxation  from static nuclear dipolar
fields in the sample holder.
Further details on
the $\mu$SR technique may be found in Ref.~\cite{musr}. 

 Fig. 1 shows several typical $\mu$SR spectra in Y$_2$Mo$_2$O$_7$.
Above T$_F$=22 K   the observed  spin 
relaxation is attributed to rapid fluctuations of the
internal magnetic field due to Mo$^{4+}$ moments 
in the paramagnetic phase. When the fluctuation rate  $\nu \gg \Delta$
 (defined below), the relaxation
function [see P$_{z}(t)$ in Fig. 1] for each magnetically equivalent muon 
site $i$ can be
described by a single exponential ${\rm e}^{-\lambda_{i}t}$ 
with a relaxation rate\cite{tomo} :
\begin{equation} 
\lambda_{i}= \frac{2\Delta_i ^2 \nu_i}{\nu_i ^2 + \nu ^2_L } 
\end{equation} 
where  $\Delta _i=\gamma _{\mu } B_i$ is
the gyromagnetic ratio of the muon ($2\pi \times 
135.54 (10^{6}$ rads s$^{-1}$ tesla$^{-1}$)) 
times the rms internal
magnetic field $B_i$ at site $i$. $\nu _i$ is the
fluctuation rate  of the internal field 
 and $\nu_L=\gamma _{\mu } B_{{\rm ext}}$ is the Larmor
frequency of the muon in the external magnetic field. Note 
that $\lambda_i$ is only weakly dependent on the   applied field 
provided $\nu_i \gg \nu_L$; this is consistent with the absence
of any field dependence observed in the spectra for T above T$_F$.
Fig. 2 shows the average muon spin relaxation rate in 
Y$_2$Mo$_2$O$_7$ obtained from fits to a single exponential 
relaxation function
P$_{z}(t)$ $\sim {\rm e}^{-t/T_1}$ 
over a restricted time interval of 0.05 to 6 $\mu$s, where $\lambda =1/T_1$.
 In the paramagnetic phase
one may use Eqn. 1 to 
estimate the average fluctuation rate of the moments.
For example with $B_i=0.066$ T (see below)  one 
obtains fluctuation rates shown in  the inset of Fig. 2. 
Note the sharp rise in  the average 
$1/T_1$ and corresponding decrease in the Mo$^{4+}$ fluctuation rate 
as one approaches T$_F$=22 K.

Just above  T$_F$,  P$_{z}(t)$ deviates somewhat from  a single exponential
(see for example T=27.5 K spectrum in Fig. 1) 
and is better described  by  a stretched
exponential of the form ${\rm e} ^{-(\lambda t)^\beta }$, with 
$\beta$  near 0.4. 
  Similar behaviour
 has  recently been observed in other  dense spin glasses 
AgMn and AuFe~\cite{campbell}. 

The muon spin  polarization function  below T$_F$ (see inset in Fig. 1) 
is characterized by rapid depolarization of
$2/3$ of the initial polarization, 
followed by  slow relaxation of the remaining $1/3$  component. 
This is a
characteristic signature of a highly disordered magnetic
state in which the  moments  are quasi-static on the timescale of the
muon lifetime. For
example,  the muon polarization function for a single
magnetic site with a Gaussian distribution of static internal
fields is  given  by the Kubo-Toyabe function~\cite{musr}: 
\begin{equation}
P_z(t)= \left[\frac{1}{3} +\frac{2}{3}(1-\Delta ^2 t^2 )
{\rm e}^{(-\frac{1}{2} \Delta ^2 t^2)}\right],
\end{equation}
The  curve in the inset of  Fig. 1 shows a fit of the
early time data at 2.5 K  to Eqn. 2,  modified slightly 
to include the small external field of 0.02 T.  
The best fit 
gives a value $\Delta /\gamma _{\mu } =0.066(3) $ T, which corresponds to 
an average field strength $\sqrt{8/\pi }\Delta /\gamma _{\mu } = 0.105(5)$ T.
 Note however that the dip in P$_{z}(t)$ at  0.032 $\mu$s is not
as deep as  predicted by the modified Eqn. 2, 
indicating the distribution of internal fields is more
complicated than a single Gaussian.  
One can generalize Eqn. 2
to include  a  fluctuating component to the internal field,
which results in relaxation of the 1/3 tail seen in Fig. 1.

Muon spin relaxation results from the exchange of energy with
 magnetic excitations.
 A first order process, in which the 
muon absorbs or creates
an excitation with an  energy equal to the muon Zeeman energy,
 is normally suppressed in conventional systems with LRO, where
 the density of states $\rho(E)\rightarrow 0$ as $E\rightarrow 0$, 
since it requires  excitations near zero energy.
In a second order (Raman magnon scattering) process involving inelastic 
scattering of an excitation, application of Fermi's Golden rule gives:
\begin{equation}
1/T_1 \propto \int _{0}^{\infty }\!\!dE\ n\left(\frac{E}{k_BT}\right)
\left[n\left(\frac{E}{k_BT}\right)
+1\right] 
M^2(E)\rho^2(E)
\end{equation}
where the muon Zeeman energy has been neglected and
 $M(E)$ is the
matrix element for inelastic scattering of an  excitation 
of energy $E$ causing a  muon spin flip. In a spin glass,
$n(E/k_BT)$ is the probability distribution (assumed to be Bose) for
``intravalley" excitations, ie. spin excitations within one of the 
macroscopic number of 
 metastable states or valleys. 
>From Eqn. 3, the
temperature dependent behaviour of $1/T_1$ is primarily determined by 
the energy dependence
of $\rho(E)M(E)$.
 The low temperature linear specific heat observed in 
Y$_2$Mo$_2$O$_7$~\cite{raju}
suggests $\rho(E)$ is flat or at least weakly dependent on energy.  
If  $\rho(E)$ and $M(E)$ have  power law dependences
with powers $l$ and $m$ respectively, 
then Eqn. 3 implies that $1/T_1$ varies as   $T^{2(l+m)+1}$ below T$_F$.
In other words, below T$_F$,  $1/T_1$ decreases gradually 
as the magnetic excitations
freeze out.
 The curve in Fig 2 shows the best  fit of the data
below 12 K to a  simple  power law form
$\lambda=\lambda _{0} + AT^n$
with exponent $n=2.1(3)$.
This power law behaviour and the
small value of $n$ indicate that $M(E)\rho(E)$ 
in Eqn. 3 has a very weak energy dependence (ie: 
$l+m$  is  less than 1).
Intervalley transitions, involving reorientations of finite sized spin
clusters, are thought to be important only in the mK 
range~\cite{tunnel}, where $1/T_1$ is independent of temperature in this
sample.
There is a small residual relaxation rate 
( $\lambda _0=$0.02 $\mu s^{-1}$ )  at the lowest temperatures, which implies 
there is a non-zero density of excitations close to zero energy.
Such relaxation is  just above the resolution limit of the 
$\mu$SR technique.

A similar spin freezing transition is observed in 
Tb$_2$Mo$_2$O$_7$, but  the 
residual $1/T_1$ at low temperatures is much larger.
Fig. 3 shows the muon spin relaxation rate measured in a small 
longitudinal field of  5 mT. As in  Y$_2$Mo$_2$O$_7$, a critical slowing 
down of the moment fluctuations occurs as one approaches T$_F$=25K from 
above. Using a value for $B_i=0.7$T
(see below) we obtain the spin fluctuation rates
($\nu$) 
above  T$_F$  shown in the inset of Fig. 3. 
For comparison, we include some of the corresponding Tb$^{3+}$ spin
fluctuation rates
determined from inelastic neutron scattering~\cite{neufluct}. 
Considering the fluctuation rates measured by neutron scattering are 
at the lower experimental limit and 
 the systematic errors in both measurements, the agreement is reasonable.
>From this we can conclude
that  both techniques are 
sensitive to the same  quantity in this sample, ie: the Tb$^{3+}$
 moment fluctuation
rates.
The fact that T$_F$ is about  the same in 
Tb$_2$Mo$_2$O$_7$ and Y$_2$Mo$_2$O$_7$
 supports the
proposal  that
the  spin freezing temperature in Tb$_2$Mo$_2$O$_7$ is determined mainly
by the Mo$^{4+}$ ions, which provide an effective coupling between the 
larger but more localized rare earth Tb$^{3+}$ moments.
%However,
%the situation in Tb$_2$Mo$_2$O$_7$ may be more 
% complicated due to the presence of the two different
%magnetic species.

Fig 3 shows that $1/T_1$ initially begins to decrease
as T falls below T$_F$ but recovers below 1 K and stays
constant at a relatively large value of 5 $\mu$s$^{-1}$. 
The initial amplitude of the relaxing
 1/3 component increases as the ratio between the external magnetic field
and internal static field. This dependence was used to estimate
 the magnitude of the static component of the internal magnetic field 
$B_i=0.70(6)$ T which is about an 
order of magnitude larger than in 
Y$_2$Mo$_2$O$_7$ ~\cite{note1}, as expected from the
ratio of Tb$^{3+}$ and Mo$^{4+}$ magnetic  moments.
This confirms  that the  Tb$^{3+}$ moments  are 
involved in the 25 K freezing transition.
Note the ratio of residual relaxation rates in Y$_2$Mo$_2$O$_7$ and 
Tb$_2$Mo$_2$O$_7$ 
is roughly equal to the ratio of the
 square of the respective  internal fields. 
 The large residual $1/T_1$ in Tb$_2$Mo$_2$O$_7$ 
establishes there is a non-zero density of low
energy excitations,  which cause relaxation either  by a first or 
second order process. 
Computer simulations by Ching {\it et al.}~\cite{ching} 
on insulating Heisenberg spin glasses Eu$_x$Sr$_{1-x}$S $(x=0.54$ and $0.40)$
 have indicated the density of states $\rho(E)$ 
may be  peaked at low energies and $\rho(0)$ finite. 
 
 We emphasize here that we find convincing evidence for a limiting
   temperature independent $1/T_1$ in Y$_2$Mo$_2$O$_7$ and
   Tb$_2$Mo$_2$O$_7$ only in the temperature range T/T$_F <0.05$.
   Previous $\mu$SR experiments~\cite{tomo,pinkvos,femn}  found a
   strong temperature dependence of $1/T_1$ in the temperature range
   T/T$_F \in [0.1-1.0]$, with no sign that $1/T_1$ was approaching a
   limiting and temperature independent value $\lim_{\rm T\rightarrow
   0}[1/T_1(T)]$ above the experimental $\mu$SR resolution limit and
   in any case, did not probe the temperature range T/T$_F<0.1$.
   It is interesting to note that other spin glasses
   like Cd$_{1-x}$Mn$_{x}$Te (0.27$\leq x \leq$ 0.65)~\cite{cdmnte} and
   La$_{1.94}$Sr$_{0.06}$CuO$_4$~\cite{lasrcuo} 
  show indications of low temperature spin dynamics but again,
 these insulating Heisenberg spin glasses have not been studied in the
   important region below 0.1T$_F$.  The geometrically frustrated
   kagom\'{e} lattice system SrCr$_8$Ga$_4$O$_{19}$ has also recently been
  studied using $\mu$SR.  Dynamics spin fluctuations are observed without
   static freezing, even at 100mK, well below T$_F$=3.5 K~\cite{kagome}.
  There is however some controversy over 
   SrCr$_8$Ga$_4$O$_{19}$~\cite{ramirez_scgo,martinez}, as it has been 
  suggested that this material does not show a thermodynamic freezing 
   transition at T$_F$~\cite{martinez}.  In this case, one would expect
  to find spin dynamics persisting down to zero temperature.
   This is not the case for Y$_2$Mo$_2$O$_7$ where we have strong
   evidence for a {\it collective freezing transition} at T$_F$ as 
   seen in the critical slowing down seen in $\mu$SR and the
 divergent nonlinear susceptibility~\cite{nonlinsus}.

In conclusion, despite the nominal absence of disorder, the freezing
process in Y$_2$Mo$_2$O$_7$ and Tb$_2$Mo$_2$O$_7$ appears  similar to
that expected for a dense spin glass.  In particular we observe  a
critical slowing down of the spin fluctuations  and non exponential
muon spin relaxation near T$_F$, while below T$_F$ there is evidence
for a highly disordered magnetic structure.  The most striking feature
in both  systems is the  presence of a residual, temperature
independent spin relaxation which persists down to very low
temperatures.  This shows there is  an appreciable  density of
states for low energy magnetic excitations which is much larger in
these systems than in conventional randomly frustrated spin glasses.
It is possible  that the residual low temperature  dynamics 
in these systems  are ``remnants'' of the zero-modes
predicted theoretically  for  nearest-neighbor Heisenberg spins on a 
pyrochlore lattice ~\cite{reimers91,harris,chalker,reimers_berlinsky}.

This research has been partially funded by the NSERC of Canada
under the NSERC Collaborative Research Grant
{\it Geometrically-Frustrated Magnetic Materials}.

\begin{figure}
\caption{ The muon spin relaxation function, P$_{z}(t)$ 
at various temperatures in Y$_2$Mo$_2$O$_7$.
The inset shows the early time behaviour at $T=2.5$ K.}
\label{fig1}
\end{figure}

\begin{figure}
\caption{  The muon spin relaxation rate $1/T_1$ vs temperature for 
Y$_2$Mo$_2$O$_7$
in a small  applied field of 0.02 T. The solid line is the best fit
to the data assuming a power law functional form.  The   Mo$^{4+}$  spin 
fluctuation rate vs temperature above T$_F$ is shown in the inset.}
\label{fig2}
\end{figure}

\begin{figure}
\caption{  Dynamical muon spin relaxation
rate $1/T_1$ vs temperature for Tb$_2$Mo$_2$O$_7$ in an applied
 field of 5 mT.  The inset shows the
 Tb$^{3+}$ moment  fluctuation rate vs temperature above 
T$_F$.  The inverted triangles indicate neutron scattering data [8].}
\label{fig3}
\end{figure}
 

\begin{references}


\bibitem{villain} J. Villain, Z. Physik B {\bf 33}, 31 (1979).

\bibitem{reimers92} J. N. Reimers, Phys. Rev. B {\bf 45}, 7287 (1992).

\bibitem{reimers91} J. N. Reimers, A. J. Berlinsky, A. -C. Shi,
 Phys. Rev. B {\bf 43}, 865 (1991).

\bibitem{harris} A.B. Harris, C. Kallin and A.J. Berlinsky, 
Phys. Rev. B {\bf 45}, 7536 (1992).

\bibitem{chalker} J. T. Chalker, P. C. W. Holdsworth and E. F. Shender,
Phys. Rev. Lett. {\bf 68}, 855 (1992).

\bibitem{reimers_berlinsky}  J. N. Reimers and A. J. Berlinsky, 
Phys. Rev. B {\bf 48}, 9539 (1991).


\bibitem{keren} A. Keren, Phys. Rev. Lett. {\bf 72}, 3254 (1994).

\bibitem{obd} S. T. Bramwell, M. J. P. Gingras, J. N. Reimers,
J. Appl. Phys. {\bf 75}, 5523 (1994).

\bibitem{fef3}  J. N. Reimers, J. E. Greedan, M. Bj\"{o}rgvinsson,
Phys. Rev. B {\bf 45}, 7295 (1992).

\bibitem{raju} N. P. Raju, E. Gmelin, R. K. Kremer, Phys. Rev. B {\bf 46},
5405 (1992).

\bibitem{nonlinsus} M. J. P. Gingras {\it et al.},
% C. V. Stager, B. D. Gaulin,
% N. P. Raju,  J. E. Greedan
 J. Appl. Phys. {\bf 79}, 6170 (1996).

%\bibitem{wandiff} J. E. Greedan, J. N. Reimers, S. L. Penny, C. V. Stager,
% J. Appl. Phys. {\bf 67}, 5967 (1990).

\bibitem{neufluct} B. D. Gaulin {\it et al.},
% J. N. Reimers, T. E. Mason, J. E. Greedan, Z. Tun,
 Phys. Rev. Lett. {\bf 69}, 3244 (1992).
 
\bibitem{musr}
%  A. Schenck, {\em Muon Spin Rotation Spectroscopy}
%(Adam Hilger Ltd, 1985);
  S. F. J. Cox, J. Phys. C {\bf 20}, 3187 (1987).

\bibitem{tomo} Y. J. Uemura {\it et al.},
% T. Yamazaki, D. R. Harshman, M. Senba, E. J. Ensaldo,
 Phys. Rev. B {\bf 31}, 546 (1985). 

\bibitem{pinkvos}  H. Pinkvos {\it et al.}, Phys Rev. B {\bf 41}, 590 (1990).

\bibitem{femn} M. J. P. Gingras {\it et al}, TRIUMF preprint (1995).

\bibitem{bandgap} M. A. Subramanian, G. Aravamudan, G. V. Subba Rao,
Mat. Res. Bull. {\bf 15}, 1401 (1980).

\bibitem{campbell} J. A. Campbell {\it et al.},
% A. Amato, F. N. Gygax, D. Herlach, 
%A. Schenck, R. Cywinski, S. H. Kilcoyne,
 Phys. Rev. Lett. {\bf 72}, 
1291 (1994).

%\bibitem{fischer} K. H. Fischer, J. A. Hertz, {\em Spin Glasses}
%(Cambridge University Press, 1991)

\bibitem{tunnel} W. Y. Ching, D. L. Huber, Phys. Rev. B {\bf 34},
1960 (1986).

\bibitem{note1}  Since the internal magnetic fields in Tb$_2$Mo$_2$O$_7$ 
are much larger
that in Y$_2$Mo$_2$O$_7$ it was not possible to resolve the fast 2/3 component 
and thereby measure $\Delta $ directly.

\bibitem{ching} W. Y. Ching, D. L. Huber, K. M. Leung, Phys. Rev. B
{\bf 21}, 3708 (1980).

\bibitem{cdmnte} E. J. Ansaldo {\it et al.},
% D. R. Noakes, J. H. Brewer, S. R. 
%Kreitzman, J. K. Furdyna, 
Phys. Rev. B {\bf 38}, 1183 (1988).

\bibitem{lasrcuo} B. J. Sternlieb {\it et al.},
% G. M. Luke, Y. J. Uemura, T. M. Riseman, 
%J. H. Brewer, P. M. Gehring, K. Yamada, Y. Hidaka, T. Murakami, T. R. Thurston,
%R. J. Birgeneau,
 Phys. Rev. B {\bf 41}, 8866 (1990). 

\bibitem{kagome} Y. J. Uemura {\it et al.},
% A. Keren, K. Kojima, L. P. Le, G. M. Luke,
%W. D. Wu, Y. Ajiro, T. Asano, Y. Kuriyama, M. Mekata, H. Kikuchi, 
%K. Kakurai,
 Phys. Rev. Lett. {\bf 73}, 3306 (1994).

\bibitem{ramirez_scgo} A. P. Ramirez, G. P. Espinosa, A. S. Cooper, 
Phys. Rev Lett. {\bf 17},  2070 (1990).

\bibitem{martinez} B. Mart\'{i}nez, A. Labarta, R. Rodr\'{i}quez-Sol\'{a},
X. Obradors, Phys. Rev B {\bf 50}, 15779 (1994).

%\bibitem{henley} J. von Delft, C. L. Henley, Phys. Rev. B {\bf 48}, 965 
%(1993).

%\bibitem{ggg} P. Schiffer {\it et al.}, Phys Rev. Lett. {\bf 74}, 2379 (1995).
\end{references}
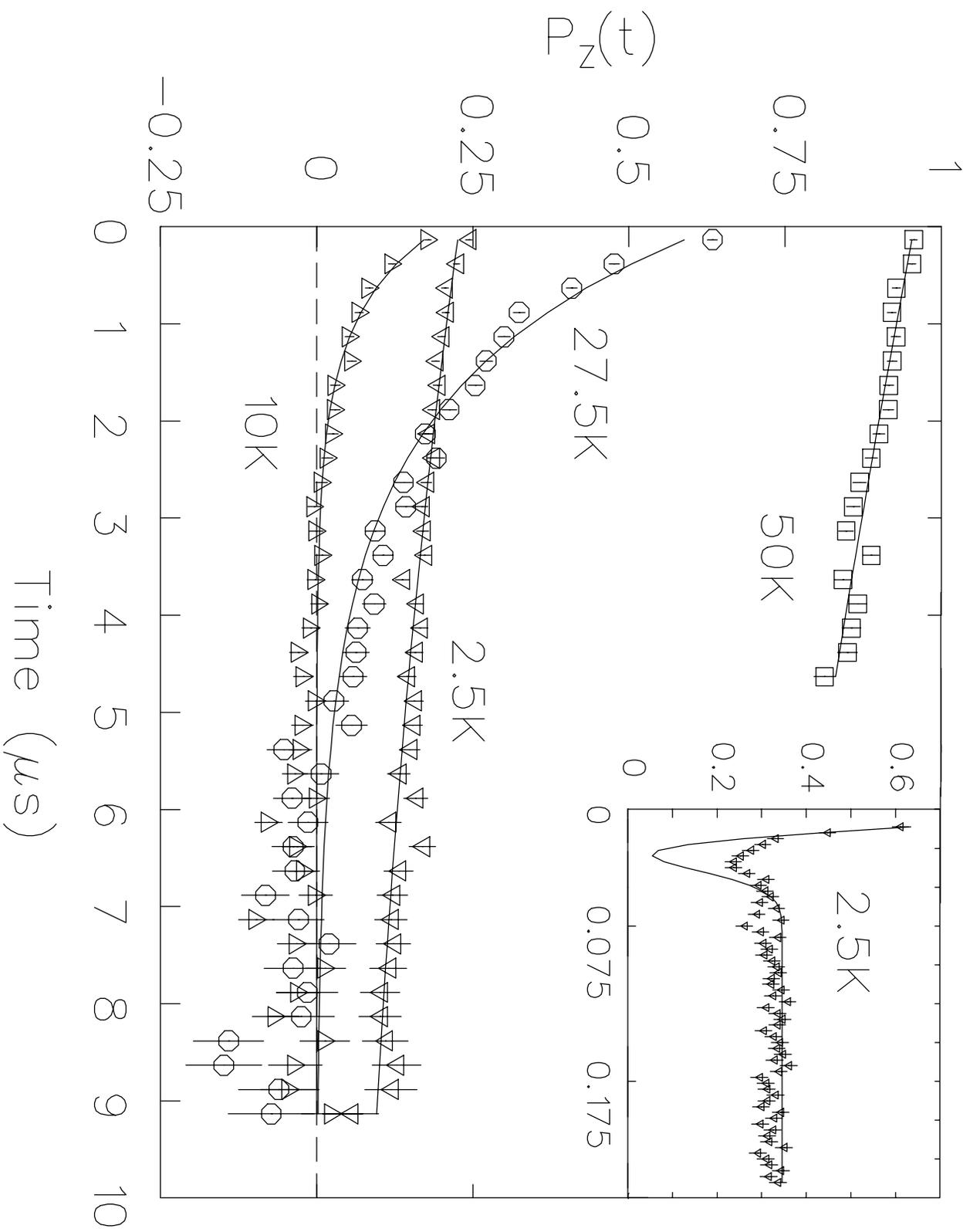
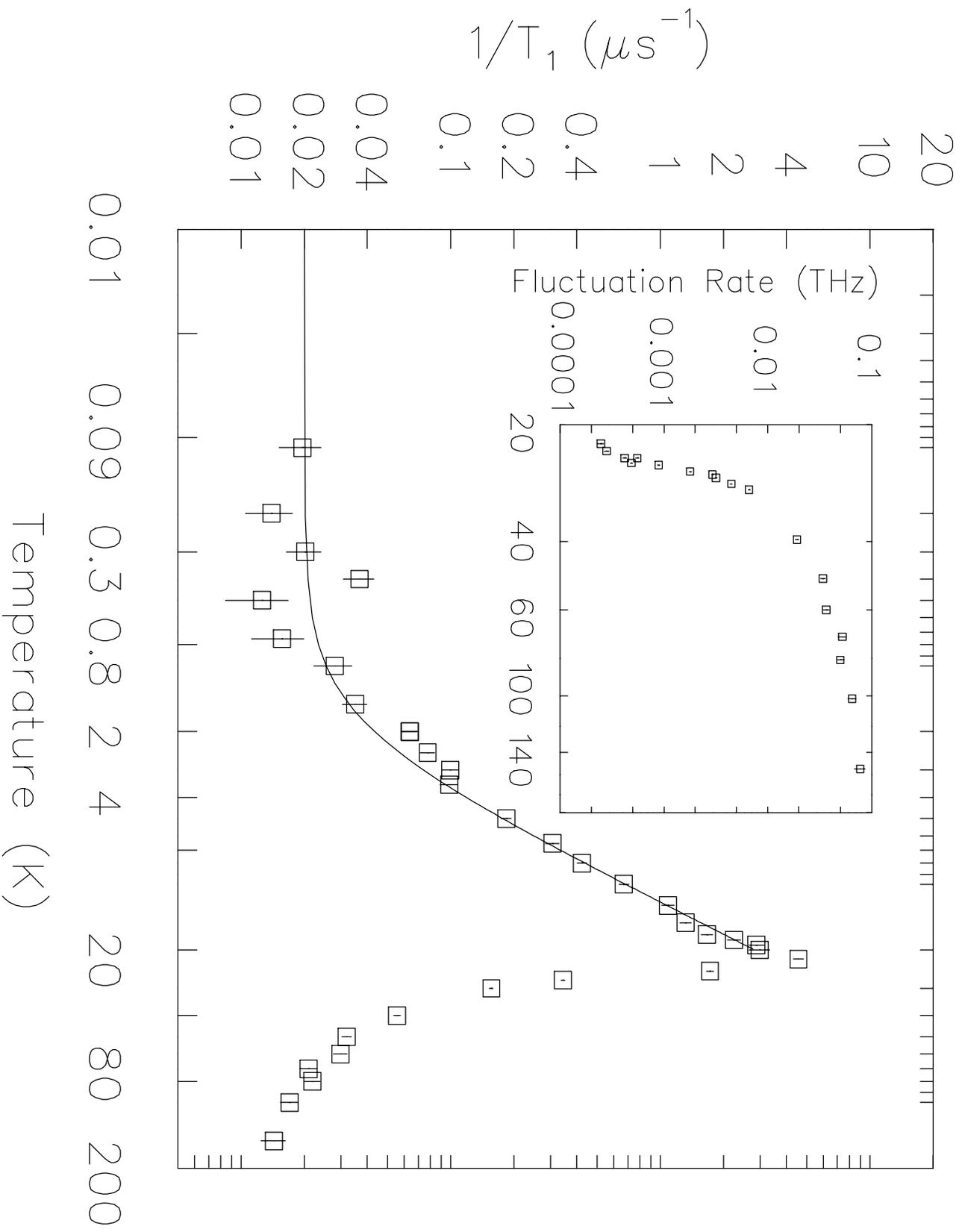
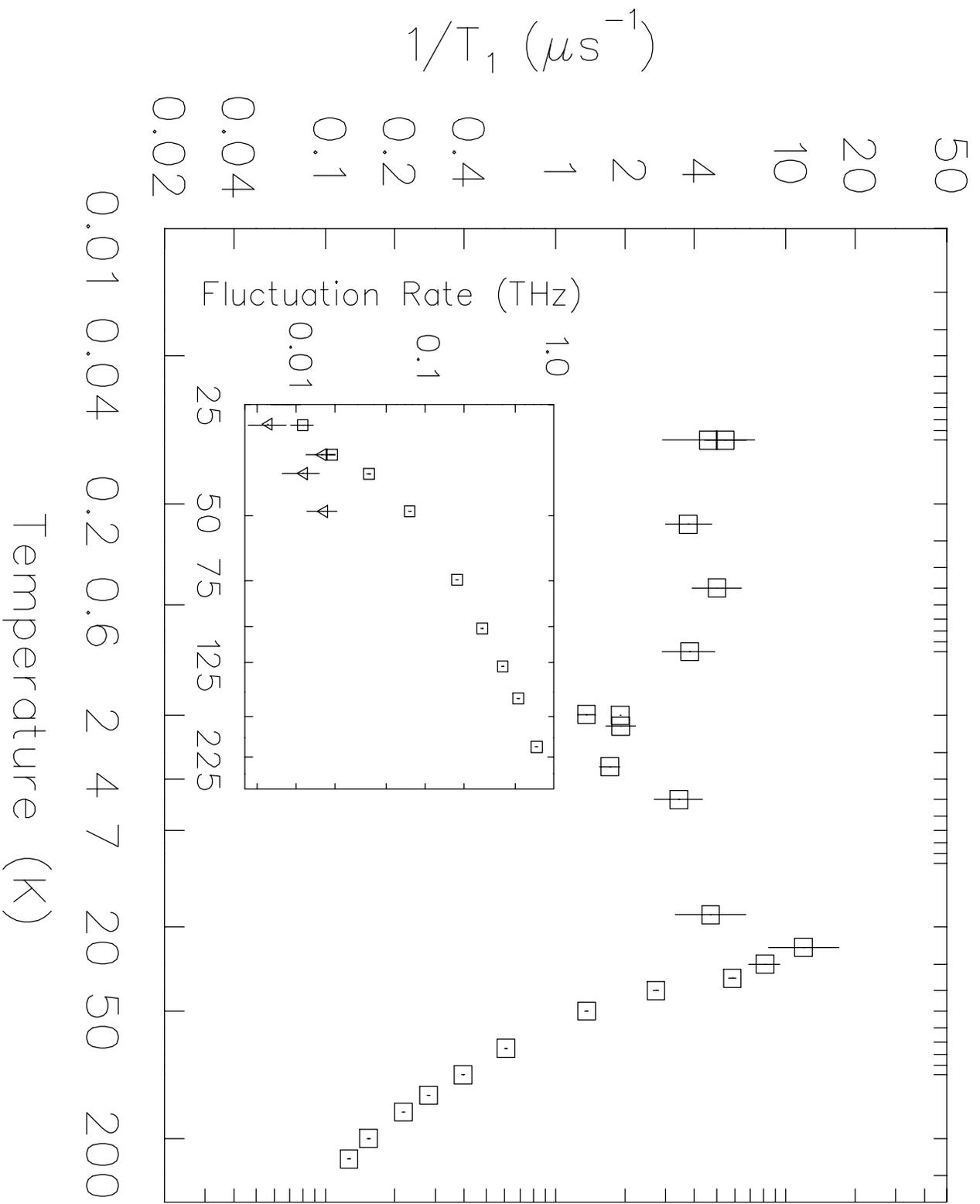
\end{document}